\documentstyle[twoside,fleqn,espcrc2,epsf]{article}

\newcommand{\AmS}{{\protect\the\textfont2
  A\kern-.1667em\lower.5ex\hbox{M}\kern-.125emS}}

\title{Toward the Continuum Limit of $B_K$ with the Quenched Kogut-Susskind 
Quark Action\thanks{presented by S. Aoki}}

\author{JLQCD Collaboration\\[2mm]
	S. Aoki\address{Institute of Physics, University of Tsukuba,
        Tsukuba, Ibaraki 305, Japan},
        M. Fukugita\address{Yukawa Institute for Theoretical Physics,
        Kyoto University, Kyoto 606, Japan},
        S. Hashimoto\address{National Laboratory for High Energy Physics (KEK),
        Tsukuba, Ibaraki 305, Japan},
        N. Ishizuka$^{\rm a}$
	Y. Iwasaki$^{\rm a,d}$,
	K. Kanaya$^{\rm a,}$\address{Center for Computational Physics, 
        University of Tsukuba, Tsukuba, Ibaraki 305, Japan},
	Y. Kuramashi$^{\rm c}$,\\
        H. Mino\address{Faculty of Engineering, Yamanashi University,
        Kofu 400, Japan},
	M. Okawa$^{\rm c}$,
	A. Ukawa$^{\rm a}$,
	T. Yoshi\'e$^{\rm a,d}$
}
       
\begin{document}

\begin{abstract}
We present new results of our ongoing project toward a precision 
determination of the kaon $B$ parameter with the Kogut-Susskind quark action 
in quenched QCD.  
New results taken at $\beta$=6.4 and $\beta=5.7$ suggest that an apparently
linear $a$ dependence of $B_K$ previously observed for $\beta=5.85-6.2$ arises
from a change of curvature from convex to concave as the lattice spacing is
reduced.  Fitting  data for $\beta\geq 5.93$ with an $O(a^2)$ form yields
$B_K$(NDR,2 GeV)=0.587(7)(17) in the continuum limit.  We also describe
a finite-size
study of $B_K$ at $\beta=$ 6.0 and 6.4, and a reanalysis of the theoretical
argument for $O(a^2)$ behavior.

\end{abstract}

\maketitle

\section{Introduction}
Lattice QCD determination of the kaon $B$ parameter with the
quenched Kogut-Susskind quark action has been pursued over a number of
years.  Last year we initiated a renewed effort toward this goal using 
VPP500/80 at KEK, improving 
upon a number of points of calculation of the previous pioneering 
studies\cite{STAG,IFMOSU,sharpe1}. In particular, with a systematic scaling
study over a wide range of lattice spacing carried out with improved
statistics, we attempted to achieve a verification of the 
$O(a^2)$ scaling violation theoretically suggested\cite{sharpe1} and then a
precision determination of the continuum value of $B_K$.  

Our initial results taken at $a^{-1}=1.3-2.6$ GeV ($\beta=5.85-6.2$), reported
at Lattice 95\cite{JLQCD95}, however, did not exhibit the
expected $O(a^2)$ behavior  but were apparently more consistent with an $O(a)$
behavior. In order to understand this result, we have 
since extended our study by a new calculation both at a larger and a 
smaller value of $a^{-1}$.  We have also carefully examined 
the question of finite-size effects in $B_K$,  
which could jeopardize the correct scaling behavior.  In addition a
reexamination of the theoretical argument for the $O(a^2)$  behavior has been
made.  Here we present results of these analyses.  

In this article we mostly concentrate on data taken with an equal mass for
$d$ and $s$ quarks.  Toward the end we briefly touch upon the question
of quenched chiral logarithms that could affect the case of unequal quark
masses. 

\section{New runs and analysis procedures}
\label{secthree}

\begin{table*}[bt]
\setlength{\tabcolsep}{0.3pc}
\newlength{\digitwidth} \settowidth{\digitwidth}{\rm 0}
\catcode`?=\active \def?{\kern\digitwidth}
\caption{Run parameters of our simulation. New runs since Lattice 95 are 
marked with ${}^*$.}
\label{tab:run}
\begin{tabular*}{\textwidth}{@{}l@{\extracolsep{\fill}}llllllllll}
\hline
$\beta$ & 5.7 & 5.85 & 5.93 &\multicolumn{3}{l}{6.0} & 6.2 &
\multicolumn{3}{l}{6.4} \\
\hline
\\[-3mm]
$L^3T$ &$12^3 24^*$ & $16^3 32$ & $20^3 40$&$24^3 64$ & $18^3 64^*$&
$32^3 48^*$&$32^3 64$ & $40^3 96^*$ & $32^3 96^*$ & $48^3 96^*$ \\
\#conf. &150 & 60 & 50 & 50 & 50 & 40 & 40 & 40 & 40 & 20 \\
skip    &1000&2000&2000&2000&&&5000&5000&&\\
$a^{-1}$(GeV)\ \ &0.806(14)& 1.36(3) & 1.59(3) &\multicolumn{3}{l}{1.88(4)} &
2.65(9) & \multicolumn{3}{l}{3.47(7)} \\
$La$(fm)&2.9 & 2.3 & 2.5 & 2.5 & 1.9 & 3.4 &2.4 &2.3 &1.8 &2.7 \\
$m_qa$ &0.02-0.08& 0.01-0.04&0.01-0.04&
\multicolumn{3}{l}{0.02-0.04}&0.005-0.02&
\multicolumn{3}{l}{0.005-0.02} \\
$m_sa/2\ \ $ &0.0563 &0.0202 & 0.0160 &\multicolumn{3}{l}{0.0126} & 0.0089
& \multicolumn{3}{l}{0.0070}\\
fit &6-16 & 10-20 & 12-26 & 20-42 & 20-42 &14-32 & 20-42&
25-69& 25-69& 25-69 \\
\hline
\end{tabular*}
\end{table*}

In Table~\ref{tab:run} we list the parameters of all our runs carried out 
so far.
Lattice sizes marked with $*$ signify new runs since Lattice 95.  
Our main sequence of runs taken at a physical spatial lattice size of 
$La \approx 2.5$fm now includes 
those at $\beta=6.4 (40^3\times 96)$ and 
at $\beta=5.7 (12^3\times 24)$.  
Finite-size studies are carried out at $\beta=6.4$, and also at $\beta=6.0$ to
examine possible lattice spacing dependence of finite-size effects.

Gauge configurations are generated with the 5-hit pseudoheatbath algorithm. 
Four values of quark masses, equally spaced in the interval given, are employed
at each $\beta$.

The method of calculation of $B_K$ is the same as reported at 
Lattice 95\cite{JLQCD95}.  We only summarize the main points here, referring to
ref.~\cite{JLQCD95} for details:
(i) The physical scale of lattice spacing is set by the $\rho$ meson
mass in the VT channel calculated on the same set of configurations.  (ii) For
the case of an equal quark mass taken for $d$ and
$s$ quark, we estimate half the strange quark mass $m_sa/2$ from
$m_K/m_\rho=0.648$.  Interpolation of data for $B_K$ to $m_sa/2$ is made by
fitting results for $B_K$ to the form predicted by chiral perturbation 
theory.  
(iii) We employ tadpole-improved operators using the quartic root of plaquette
as the mean link $u_0=P^{1/4}$.  The one-loop renormalization factors, which
include finite parts for the continuum $\overline{MS}$ scheme with the naive
dimensional regularization (NDR), are evaluated with
$g^2_{\overline{MS}}(1/a)$ evolved from $g^2_{\overline{MS}}(\pi/a)=P/g^2_0+
0.02461$. (This point differs from the choice $g^2_{\overline{MS}}(\pi/a)$ 
adopted at the time of  Lattice 95\cite{JLQCD95}.)  Setting $\mu=2$GeV in the
logarithm of the renormalization factors, we extract $B_K$ in the
continuum $\overline{MS}$ scheme with NDR, which we denote as $B_K($NDR, 2GeV).
(iv) We estimate errors by a single elimination jackknife procedure.

\begin{figure}[bt]
\centerline{\epsfxsize=7.5cm \epsfbox{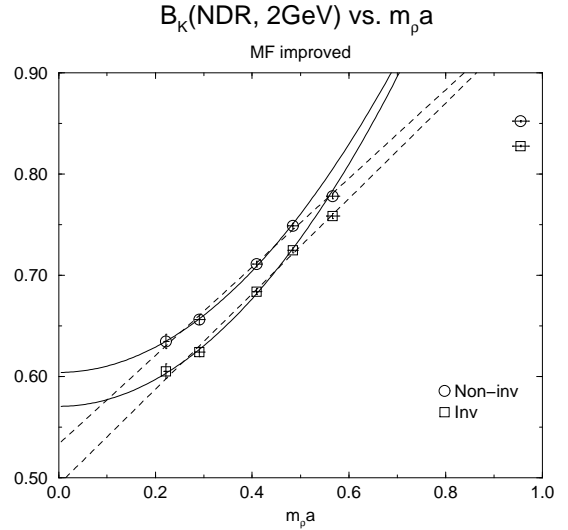}}
\vspace*{-12mm}
\caption{Gauge non-invariant(circles) and invariant(squares) 
$B_K$({\rm NDR}, 2GeV) as a function of 
$m_\rho a$, together with a quadratic fit to four data points on the left 
(solid lines) and a linear fit to five points (dashed lines).}
\label{bkmro}
\vspace*{-5mm}
\end{figure}

\section{Lattice spacing dependence of $B_K$}

In Fig.~\ref{bkmro} we present $B_K$(NDR, 2 GeV) from our main sequence of 
runs as a function of
$m_\rho a$.  The left and rightmost points are the new data taken at
$\beta=6.4$ on a $40^3\times 96$ lattice and at 
$\beta=5.7$ on a $12^3\times 24$ lattice.  

As we observed at the time of Lattice 95, the middle
four points are consistent with a linear $a$ dependence.  Viewed more closely, 
the fourth point on the right lies below, while the other three are on a 
straight line.  This suggests a convex shape of the curve of
$B_K$ toward larger lattice spacings, which is confirmed by the new data at
$\beta=5.7$ (the rightmost point in Fig.~\ref{bkmro}). 

On the other hand, the new data at $\beta=6.4$ at the
leftmost lies higher than the straight line, which is suggestive of an onset 
of an $O(a^2)$ behavior for smaller lattice spacings.  The solid line 
shows a fit of form $B_K=c_0+c_1(m_\rho a)^2$ to
the four points, excluding the two points on the right at $\beta=5.85$ and 5.7.

We note, however, that the data by themselves still do not
allow a distinction between an $O(a)$ and $O(a^2)$ dependence;
five points for $m_\rho a\leq 0.6$ can also be fitted with the form
$B_K=c_0+c_1m_\rho a$ (dashed lines).

Let us also note that the discrepancy of values between gauge invariant and
non-invariant operators, even in the continuum limit,
represents a systematic uncertainty of
$O(g^4)$
arising from the renormalization factors which are correctly
incorporated only to the order of one loop.
 
\begin{figure}[tb]
\centerline{\epsfxsize=7.5cm \epsfbox{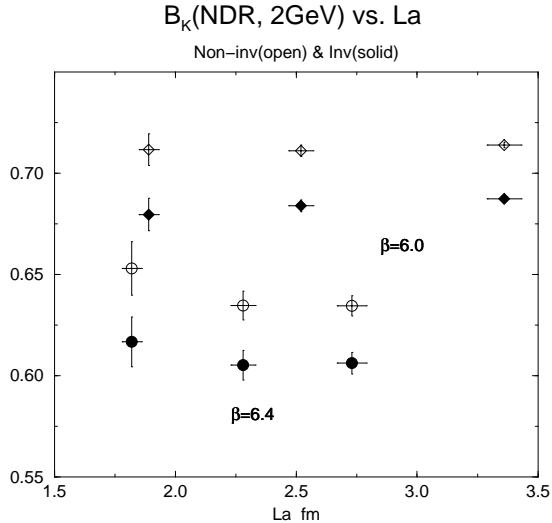}}
\vspace*{-12mm}
\caption{$B_K$({\rm NDR}, 2GeV) as a function of physical spatial size 
$L a$ (fm) at $\beta = 6.0$ and $6.4$.}
\label{BKLa}
\vspace*{-5mm}
\end{figure}

\section{Lattice size dependence of $B_K$}

We should emphasize that a distinction between an $O(a)$ or $O(a^2)$
dependence rests on a very tiny effect in the range of lattice spacing being
examined; a 2\% decrease of the point at
$\beta=6.4$ would render the $O(a)$ behavior strongly favored.  Clearly a
reexamination of finite-size effects is needed, especially since our
physical spatial size decreases by about 10\% from 
$La=2.5$fm to 2.3fm between $\beta=6.0$ and 6.4.

In Fig.~\ref{BKLa} we summarize results of our finite-size study at 
$\beta=6.0$ and 6.4, plotting $B_K$(NDR, 2GeV) as a function of physical
spatial size $La$.  For $La \geq 2.3$fm, we do not observe a 
size-dependent
shift of $B_K$ beyond the statistical error of at most 1\%.  Therefore, we
conclude that the data in Fig.~\ref{bkmro}, taken with $La$=2.3-2.9 fm, are not
affected by finite size  effects.  In particular, the rise of $B_K$ at
$\beta=6.4$  is not a finite size effect.

An unexpected feature in Fig.~\ref{BKLa} is that finite-size effects below
$La\approx 2$fm is markedly enhanced at a weaker coupling of $\beta=6.4$ than at
6.0.  Origin of this behavior is not clear to us.

\section{Reexamination of the theoretical argument for $O(a^2)$ behavior}

A subtle point in the theoretical analysis of $a$ dependence of $B_K$ is 
that numerical simulations employ 4-component hypercubic fields
$q(y),\overline{q}(y)$ defined on a coarse lattice of lattice spacing $2a$, in
terms of which the Kogut-Susskind quark action has an apparent
$O(a)$ term:
\begin{eqnarray}
S_{KS}
&=& \sum_{y}
    \bar{q}(y) [   ( \gamma_\mu \otimes I ) D_\mu + m ( I \otimes I) \cr
&&                + a ( \gamma_5   \otimes \xi_\mu \xi_5 ) D_\mu^2 
                    + a {i g} F_{\mu\nu} T_{\mu\nu} /4 \cr
&&                 + O(a^2) ] q(y) .
\nonumber
\end{eqnarray}
However, a redefinition of fields given by
\[
Q=( 1 - \sum_\mu P_\mu D_\mu /2 ) q ,    
\bar{Q} = \bar{q}( 1 - \sum_\mu P_\mu \stackrel{\leftarrow}{D}_\mu /2)
\nonumber
\]
with $ P_\mu =(1-\gamma_\mu\gamma_5\otimes \xi_\mu \xi_5)/2$ removes the $O(a)$
terms at tree-level so that the action becomes
\begin{eqnarray}
S_{KS}&=& \sum_y
     \bar{Q}(y) [  ( \gamma_\mu \otimes I ) D_\mu + m ( I \otimes I)
                 ] Q(y)\nonumber\\
&&+O(a^2).
\nonumber
\end{eqnarray}
For bilinear and four-quark operators it is also possible to show that
\begin{eqnarray}
&&\langle P | {\cal O}( \bar{Q} Q ) | P' \rangle \cr
&=& \big[ 1 + \sum_\mu i ( {P'}_\mu - P_\mu )/4 \big]
    \langle P | {\cal O}( \bar{q} q ) | P' \rangle \cr
&&+ O(a^2) . 
\nonumber
\end{eqnarray}
Therefore $B_K$ evaluated with $q(y)$'s with $P=P'$ 
has no $O(a)$ corrections at tree level.

The rest of the argument essentially follow that of ref.~\cite{sharpe1}.
In the presence of interactions
the action expressed in terms of $Q$ fields may contain $O(a)$ terms.  
However, using symmetry of the original action under 
translation, 
$Q(y) \rightarrow (I\otimes\xi_\mu) ( 1 + \partial_\mu/2 )Q(y')$,
rotation, 
$Q(y) \rightarrow ((1-\gamma_\mu\gamma_\nu)
\otimes(1-\xi_\mu\xi_\nu))/2 Q(y')$, 
and $U(1)_A$ symmetry,
$Q(y) \rightarrow {\rm e}^{i\alpha (\gamma_5\otimes\xi_5)} Q(y)$ performed
simultaneously with a mass rotation, 
$m \rightarrow {\rm e}^{-i\alpha (\gamma_5\otimes\xi_5)} m 
{\rm e}^{-i\alpha (\gamma_5\otimes\xi_5)}$,
one can show that
there are no dimension five operators giving rise to an $O(a)$ term. 
Therefore, the original action is itself improved up to $O(a^2)$.
One can also check that the $B_K$ operator is similarly improved to $O(a^2)$
since there are no dimension seven operators with the same symmetry 
properties as 
the original one.

\section{Extrapolation to the continuum limit}

Our theoretical reexamination supports the original assertion\cite{sharpe1} 
that
there should be no $O(a)$ corrections in  $B_K$  including gauge interactions.
This leads us to attempt a continuum extrapolation of our data assuming the 
$O(a^2)$ dependence.  

Fitting data in Fig.~\ref{bkmro} to the purely quadratic form
$B_K=c_0+c_1(m_\rho a)^2$, excluding the two points on the right at
$\beta=5.85$ and 5.7, we find 
\[
B_K({\rm NDR, 2 GeV}) = 0.587 \pm 0.007 \pm 0.017
\nonumber
\]
where the first error is statistical and the second  is
the systematic error of $O(g^4)$
estimated from the difference of extrapolated values for  
gauge non-invariant and invariant operators.
Changing the number of points and/or including higher order terms
in $m_\rho a$, we find that  the extrapolated
values stay consistent with the above value.

\section{$B_K$ with unequal quark masses}

We measure $B_K$ also for unequal quark masses for $d$ and $s$ quark in our
simulation.  In this case, a potential problem of quenched chiral logarithm 
may arise in the extrapolation to the chiral limit in $m_d$.

We examine this problem through
fits of our data for $m_d\ne m_s$ to the form predicted by quenched chiral
perturbation theory\cite{sharpe2}.  Generally we find very good fits.  However,
the parameter
$\delta=m_0^2/48\pi^2 f_\pi^2$ multiplying the quenched chiral logarithms turns
out negative albeit with a large error, $\delta\approx -0.3\pm 0.3$ (we note
that the sign of the $\delta$ term in ref.~\cite{sharpe2} is
incorrect\cite{sharpe3}).  
Also fits excluding the logarithm are quite acceptable, only
slightly worse than those with it. Thus we cannot conclusively conclude 
the presence of quenched chiral logarithms.

Let us add the remark that the mass inside the quenched chiral logarithm should
theoretically be that of the non-Nambu-Goldstone pion\cite{sharpe3}.  Fits in
this case differ from those with the Nambu-Goldstone pion mass only very close to
$m_d=0$, however, so that the conclusion above does not change.

\section{Conclusions}

Our new data for $B_K$ extending the results of last year to smaller and larger
lattice spacings suggest that the curve of $B_K$ changes curvature from
convex to concave as the lattice spacing is reduced from 
$m_\rho a\approx 0.9$ to 0.2.  This behavior
explains an apparently linear dependence we observed in our previous data, and
is also consistent with an $O(a^2)$ dependence theoretically expected toward
the continuum limit.  To confirm if $B_K$ stays on an $O(a^2)$ curve, 
we plan to carry out one more run on a
$56^3\times 96$ lattice at $\beta$=6.65 with yet smaller lattice spacing of
$m_\rho a\approx 0.16$.

\end{document}